\documentclass[12pt]{article}
\usepackage{amsfonts}
\usepackage{amsmath}
\usepackage{amssymb}
\usepackage{amsthm}
\usepackage{bigints}
\usepackage[font=small,labelfont=bf]{caption}
\usepackage{color}
\usepackage{csquotes}
\usepackage{epstopdf}
\usepackage{float}
\usepackage{footmisc}
\usepackage{graphicx,datetime}
\usepackage{geometry}
\usepackage{hyperref}
\usepackage{hypcap}
\usepackage{mathtools}
\usepackage{natbib}
\usepackage{pdfpages}
\usepackage{pgfplots}
\usepackage{rotating}
\usepackage{setspace}
\usepackage{subfig}
\usepackage{tikz}
\usetikzlibrary{calc}
\usetikzlibrary{patterns}
\usepackage{xcolor}

\usetikzlibrary{calc}

\newtheorem{lemma}{Lemma}

\newtheorem{proposition}{Proposition}

\hypersetup{
colorlinks,
linkcolor={red!50!black},
citecolor={blue!50!black},
urlcolor={blue!80!black}
}

\begin{document}

\title{Nonlinear Prices, Homogeneous Goods, Search\thanks{\small%
		\ I thank Maarten Janssen, Karl Schlag, and Nicolas 
		Schutz for valuable comments.}}

\author{Atabek Atayev \thanks{\small Corresponding author; 
		e-mail: atabek.atayev@zew.de}\\
	{\small ZEW---Leibniz Centre for European Economic Research 
	in Mannheim}\\
	{\small L 7, 1, 68161 Mannheim, Germany}}%
\date{\today\\
}

\begin{singlespace}
	\maketitle
		
\end{singlespace}
\begin{abstract}
	\noindent 
	We analyze competition on nonlinear prices in homogeneous 
	goods markets with consumer search.  In equilibrium firms 
	offer two-part tariffs consisting of a linear price and 
	lump-sum fee.  The equilibrium production is socially 
	efficient as the linear price of equilibrium two-part 
	tariffs equals to the production marginal cost.  Firms thus 
	compete in lump-sum fees, which are dispersed in 
	equilibrium.  We show that sellers enjoy higher 
	profit, whereas consumers are worse-off with two-part 
	tariffs than with linear prices.  The competition softens  
	because with two-part tariffs firms can make effective 
	per-consumer demand less elastic than the actual demand.
	
	\bigskip
	
	
\end{abstract}

\sloppy

\newpage

\section{Introduction}\label{sec:intro}

In this paper we analyze nonlinear prices in homogeneous goods 
market with imperfect competition.  Nonlinear prices are 
widespread. Examples include markets for utilities, consumer 
financial credits, and telecommunication services. In these 
markets goods are fairly homogeneous across sellers, while 
buyers are usually offered some form of nonlinear prices.  For 
instance, German electricity providers offer two-part tariffs 
which consists of a linear price and a monthly lump-sum fee. A 
total price a buyer pays for a credit consists of interest 
payments and a fixed service fee.   A mobile-phone buyer's total 
payment includes a monthly fee for a fixed data and a lump-sump 
fee at a time of signing a contract.  

Despite prevalence of nonlinear prices in competitive 
homogeneous goods markets, models of such markets remain 
understudied. The reason for this is the argument that, 
in markets where firms compete on prices, equilibrium offers do 
not depend on the type of pricing scheme firms employ.  This is 
because sellers do not possess any market power.  If a seller 
offers a slightly more expensive deal than its rivals, it loses 
all its buyers to the competitors. Therefore independent of the 
pricing scheme, production is socially optimal and consumers 
receive the entire social surplus. 

In real-world markets, however, sellers exercise market power 
even in homogeneous goods markets as some consumers cannot 
easily walk away to competing sellers when these consumers are 
offered a bad deal.  The reason for this is that inspecting 
sellers' offers entails a significant time cost.  For some 
buyers this cost may be so high that they prefer foregoing 
search and buying from a current seller at a high price (see 
e.g., \cite{delossantosetal2012}, \cite{kaplanmenzio2015}, 
\cite{alexandrovkoulayev2017}, \cite{lachmoragagonzalez2017}, 
\cite{gorodichenkoetal2018}, \cite{gugleretal2018}, 
\cite{galenianosgavazza2020}).  While sellers possess market 
power over these buyers, sellers also have incentive to compete 
for consumers who can easily walk away to competitors.  
These two opposing incentives causes imperfect competition which 
oftentimes results in price dispersion (e.g., \cite{varian1980}, 
\cite{burdettjudd1983}, \cite{stahl1989}). Although there is a 
large body of literature studying consumer search markets, they 
are typically limited to linear prices.  There are a few reasons 
for this.  First, linear prices are easy to examine and thus 
they serve as a natural first step in analyzing search markets.  
Also in markets with indivisible goods and unit 
demand, equilibrium outcomes with certain types nonlinear 
prices, such as two-part tariffs, coincide with that of linear 
prices.

%These models suit to analyze markets where buyers demand a unit 
%of the product such as a car or an insurance.  In such markets, 
%a total price a buyer pays for a unit of the product may 
%consists of different fees. Nevertheless, this total price is a 
%price per unit, meaning that nonlinear prices are 
%inconsequential in these markets.%
%\footnote{Nonlinear prices may be employed in a homogeneous 
%goods market with a unit demand if firms have additional 
%information about buyers such as their search history }

In this paper we employ traditional models of sequential 
consumer search with homogeneous goods to study nonlinear 
prices.  Standard features of such models are as follows. 
Oligopoly firms supply homogeneous goods to a large number of 
buyers and compete on prices.  Consumers have downward sloping 
demand and differ in their costs of learning firms' offers.  
When facing a same offer, a buyer with a low search cost is  
likely to sample an offer of another firm whereas a buyer with a 
high search cost is likely to forego search and buy outright.  
The novel feature of our model is that firms can offer nonlinear 
prices.

We show that firms compete on two-part tariffs in equilibrium.  
A two-part tariff consists of a linear price and a fixed fee.  
The linear price of any equilibrium two-part tariff is equal to 
the production marginal cost.  Therefore production is socially 
optimal with nonlinear prices. To contrast, if firms can offer 
linear prices only, the equilibrium prices (which are dispersed) 
are higher than the production marginal cost 
(\cite{stahl1989}).  Thus the social surplus with linear prices 
is lower than that with nonlinear prices.   

Fixed fees of equilibrium two-part tariffs are dispersed, 
meaning that sellers compete in fixed fees to attract buyers 
with low search costs. The equilibrium profit is higher and 
consumer surplus is lower with two-part tariffs than with linear 
prices.  To understand the intuition one can think of firms 
offering two-part tariffs as them offering a unit of a 
hypothetical product, which contains socially optimal amount of 
the real product, at a fixed fee.  Hence, sellers effectively 
face inelastic demand per-buyer.  In contrast, if firms can only 
compete in linear prices, they set prices at the elastic part of 
the demand function. As sellers exercise greater market power 
with inelastic demand than with elastic one, their profits are 
higher with nonlinear prices than with linear prices. This in 
turn harms buyers.

Our paper contributes to two stands of the literature.  One is 
on nonlinear prices.  Early studies on nonlinear prices focus 
predominantly  on markets with a monopolist seller (e.g., 
\cite{oi1971}, \cite{spence1977}, \cite{mussarosen1978}, 
\cite{stokey1979}).  Later studies extend these models to 
competitive environment, yet their analysis centers on markets 
with differentiated products (e.g., \cite{hayes1987}, 
\cite{armstrongvickers2001}, \cite{rochetstole2002}, 
\cite{yin2004}, \cite{garrettetal2019}, \cite{tamayotan2021}).  
In some of these studies it is possible to interpret qualities 
that firms offer as quantities offered.  Doing so one would 
reformulate a market of vertically differentiated products as 
a homogeneous goods market where firms compete in quantities 
offered, i.e., they compete a la Cournot.  This would deviate 
from our study as we consider competition on prices.

The other branch of literature that is close to our paper is the 
literature on consumer search.  Traditional studies of consumer 
search with homogeneous product and linear prices include 
\cite{salopstiglitz1977}, \cite{varian1980}, 
\cite{burdettjudd1983}, and \cite{stahl1989}.  
\cite{shelegiawilson2021} is a recent paper that touches the 
issue of nonlinear prices in markets where buyers have imperfect 
information on firms' offers.  The authors provide a useful 
method to establish equilibrium with two-part tariffs in a model 
where consumers differ in the number of offers they observe (as 
in \cite{butters1977} and \cite{varian1980}).  One of the main 
differences of their study is that buyers do not really search, 
i.e., they passively wait for advertisements to arrive and then 
decide from which firm to buy goods.%
\footnote{In the appendix of their paper, 
\cite{shelegiawilson2021} consider an extension of their main 
model where buyers who received an ad can search one more firm.}
In our paper buyers can actively search for different 
deals.  Also whereas the contribution of their paper is largely 
methodological, we are interested in welfare implications of 
nonlinear prices. To the best of our knowledge 
\cite{janssenparakhonyak2013} is the only paper that studies a 
form of nonlinear prices in homogeneous goods markets with 
search and undertakes welfare analysis.  The authors consider 
a market where a firm can match its price to a 
competitor's price if a buyer credibly demonstrates that the 
competing firm charges a price lower than the price of the firm 
under question.  This form of nonlinear price is optimal only if 
consumers with high search costs can somehow observe prices of 
multiple sellers without searching.  In our paper a consumer has 
to actively search to observe a firm's deal.

The rest of the paper is organized as follows.  In the section 
we lay out out main model.  In sections \ref{s:eq} and 
\eqref{s:welfare} we provide equilibrium analysis and welfare 
analysis, respectively.  In section \ref{s:search} we extend the 
main model to allow for a more general distribution of search 
costs and for a mixture of simultaneous and sequential search. 
We show that our main result is robust to such extensions.  The 
final section concludes.

\section{Model}\label{s:model}

There are $n$ identical firms that produce homogeneous goods at 
a constant marginal cost normalized to zero.  Firms compete in 
prices.  A unit mass of buyers, or consumers, wish to consume 
the product. Each consumer is willing to buy $q(p)$ amount of 
the product at price $p$.  Before searching, consumers do not 
know offers of firms and need to engage in sequential search to 
learn them.  To avoid \textit{Diamond-paradox} equilibrium where 
consumers do not search beyond the first firm 
(\cite{diamond1971}), we assume that a positive share of buyers 
observe offers of multiple firms.%
\footnote{This is perhaps the most common way of overcoming
	Diamond-paradox in consumer search literature (e.g., 
	\cite{salopstiglitz1977}, \cite{varian1980}, 
	\cite{stahl1989}).}
Specifically, we say that $\lambda \in (0,1)$ share of buyers 
have zero search cost and thus they observe all prices in the 
market.  The remaining share of buyers incur a positive search 
cost, denoted by $s>0$, each time they search a firm.  Following 
a terminology proposed by \cite{stahl1989} we refer to consumers 
with zero search cost as \textit{shoppers} and those with a 
positive search cost as \textit{nonshoppers}.  Identity of each 
consumer is her private knowledge. To ensure that there is some 
search we assume that searching the first firm is free.

%Notice that if the population of consumers consists of only 
%shoppers, the unique equilibrium is \textit{Bertrand-paradox} 
%where firms price at the production marginal cost.  In contrast 
%if the nonshoppers are the only buyers, the unique equilibrium 
%is \textit{Diamond-paradox} where buyers do not search beyond 
%the first firm and, thus, firms have monopoly power 
%(\cite{diamond1971}).  Presence of both types of consumers rule 
%out these extreme market outcomes.

The demand function $q(p)$ is non-increasing in $p$ with
$q(\overline{p})=0$ for some finite $\overline{p}>0$.  Letting 
$\pi(p)\equiv q(p)p$ be the revenue per consumer that price $p$ 
generates, we assume that there exists $p^m$ that maximizes 
$\pi(p)$.  This assumption is justified if the demand elasticity 
is increasing in price:
\begin{equation}\label{as:pm}
	-\frac{q'(p)p}{q(p)}\  \mbox{strictly 
	increases with} \ p.
\end{equation}
We let $\pi^m \equiv \pi(p^m)$ to simplify the notation. 

It is useful to define a function $v(\pi)$ to represent a 
surplus a buyer enjoys when generating revenue $\pi$.  This 
function is decreasing in $\pi$ as per-consumer revenue is 
increasing in price on $[0,p^m]$ and consumer surplus 
$\int^{\overline{p}}_{p}q(z)dz$ decreases with 
$p$.  Assumption in \eqref{as:pm} implies that $v(\pi)$ is 
strictly concave on $[0,\pi^m]$, which can be seen 
differentiating $v(\pi(p))$  to obtain
\begin{equation}
	v'(\pi(p)) = -\dfrac{1}{1 + \dfrac{q'(p)p}{q(p)}}.
\end{equation}

The timing of the game is as follows. First, firms 
simultaneously choose their sales strategies.  Second, without 
knowing firms' offers, consumers engage in search.  Having 
observed an offer of at least one firm, consumers may make 
purchases.  

The solution concept is Reservation price equilibrium (RPE).  
An RPE is a perfect Bayesian equilibrium where buyers search 
according to the reservation price rule.  If in a model where 
firms charge linear price a buyer searched $k$ number of firms, 
her reservation price is the lowest of those $k$ prices such 
that she is indifferent between buying at that price and 
searching further. If the lowest price a buyer has observed is 
lower than the reservation price, she buyers outright.  If this 
price is higher than the reservation price, she searches 
further.  It is typical to assume that buyers have passive 
off-equilibrium beliefs.  It means that if a buyer observes a 
deal which is not a part of equilibrium, she believes that 
sellers, whom she has not searched yet, play equilibrium 
strategy.   In a model where firms employ nonlinear prices, we 
need to replace a linear price with a total price and the above 
definition of equilibrium goes through.  We focus on symmetric 
equilibria.

\section{Equilibrium}\label{s:eq}

We first consider a case where firms can offer nonlinear 
prices and then a case where firms compete in linear prices.  
For each case, the analysis consists of two parts.  Assuming 
that there is an equilibrium in nonlinear prices, we 
characterize this equilibrium.  We later verify that the 
equilibrium indeed exists.

\subsection{Nonlinear Prices}

We start the analysis by making some important observations.  
First, it is clear that firms never offer nonlinear prices that 
yield a negative per-consumer profit.  It is also clear that it 
is sub-optimal to offer a deal that renders negative utility to 
buyers, as buyers would never accept such a deal and, thus, a 
firm offering this deal does not make any sales.

To make further observations it is useful to think of firms as
offering utilities to buyers.  Our next observation is that a 
symmetric equilibrium strategy of firms must be in 
mixed-strategies.  The reasoning can be established by 
contradiction.  Suppose that there exists a symmetric 
equilibrium in a pure strategy.  Due to symmetry, all firms 
must offer the same level of utility to buyers in equilibrium 
and therefore serve equal share of nonshoppers as well as 
shoppers.  However, an individual firm has an incentive to offer 
a slightly higher utility to attract all shoppers and increase 
its profit, a contradiction. 

We generalize this result in the following lemma.

\begin{lemma}\label{lem:no_atom}
	If there exists a symmetric equilibrium with nonlinear 
	prices, it must be in mixed-strategies and an equilibrium 
	distribution of utilities has no atom.
\end{lemma}

The understanding why firms do not offer any utility with a 
strictly positive probability is similar to the idea why there 
does not exist a symmetric equilibrium in pure strategies.  
Suppose for contradiction that a (symmetric) equilibrium 
distribution of utilities offered contains an atom at some 
utility level.  This means that all firms tie at that utility 
level with a strictly positive probability, in which case they 
serve equal share of buyers.  This cannot be optimal as an 
individual firm has an incentive to offer a slightly higher 
utility to attracts all shoppers and, thus, discontinuously 
raise its profit.  Thus we arrive at a contradiction.

For our next observation we define a reservation utility of 
a nonshopper at a given search round as the lowest utility at 
which she buys outright foregoing search.  In a symmetric 
equilibrium a nonshopper samples utility from the same utility 
distribution in each search round.  In this case it is 
well-established that the reservation utility of a nonshopper is 
stationary (\cite{kohnshavell1974}), i.e., it is the same at 
each search round.  Clearly, if a nonshopper searches all firms 
she buys at a firm offering the highest (positive) utility 
level.  If the highest utility level offered is negative she 
does not make a purchase.  This bring us to the next observation 
which we state in the next lemma.

\begin{lemma}\label{lem:reserv_utility}
	If there exists a symmetric equilibrium with nonlinear 
	prices and a positive reservation utility, no firm offers a 
	utility below the reservation utility.
\end{lemma}

The reasoning is by contradiction. Assume that some, or all, 
firms offer utility levels lower than the reservation utility.  
All consumers who happen to visit such a firm search further.  
Therefore this firm makes positive sales only if all the other 
firms happen to offer a utility lower than the utility of the 
firm under question.  If, however, all firms offer such low 
utilities, it is optimal to offer the highest of those 
utilities, which follows from a standard Bertrand-type of 
argument.  This highest utility must be equal to the reservation 
utility since there is no atom in the utility distribution, a 
contradiction.

We are now ready to make the final observation about firms' 
strategies. 

\begin{lemma}\label{lem:no_gap}
	If there exists a symmetric equilibrium with nonlinear 
	prices and a positive reservation utility, cumulative 
	distribution of utilities does not have a flat region in the 
	support.
\end{lemma}

The reasoning is by contradiction, and therefore assume that an 
equilibrium distribution of utilities has a flat region in the 
support.  A firm expects to sell to the same share of buyers 
both at the highest utility level and at the lowest utility 
level in that flat region.  In equilibrium an individual firm 
must be indifferent between offering those two utility levels. 
This can be the case only if per-consumer revenues generated by 
these utility levels are the same.  These per-consumer revenues 
are the same only if the social surplus generated by offering 
the low utility level is lower than that generated by offering 
the high utility level.  However, a firm is better-off if it 
chooses the high utility level but charges an additional 
lump-sum fee so that the final utility offered is the same as 
the low utility level. This leads to a contradiction. 

The above three lemmas imply that, given a reservation utility 
of nonshoppers, the distribution of utilities is continuously 
increasing in its compact support.  We use this fact to 
determine optimal (nonlinear) pricing strategy that results in 
such distribution of utilities.  Letting $(p,t)$ represent a 
two-part tariff where $p$ is the linear price and $t$ is a 
lump-sum fee, we state our next result in the following lemma.

\begin{lemma}\label{lem:p=0}
	If there exists a symmetric equilibrium in nonlinear prices, 
	firms compete in two-part tariffs $(p,t)$ where $p=0$.
\end{lemma}

The reasoning is as follows.  Let utility level $\widetilde{u}$ 
be in the support of the utility distribution conditional on the 
reservation utility.   If a nonlinear price that results in an 
offer of this utility level also generates a social surplus 
given by $S$, the associated per-consumer revenue equals to $S - 
\widetilde{u}$.  A firm offering utility $\widetilde{u}$ then 
chooses a nonlinear price that maximizes the social surplus. It 
is easy to see that the social surplus is maximized at a linear 
price equal to the production marginal cost.  Then, a two-part 
tariff is a part of equilibrium if its linear price equals to 
the production marginal cost and its lump-sum fee equals to 
$v(0) - \widetilde{u}$, where $v(0) = 
\int_{0}^{\overline{p}}q(z)dz$ is the highest social surplus.  

Lemma \ref{lem:p=0} allows us to reformulate firms' problem in 
terms of two-part tariffs.  Importantly, the lemma informs us 
that firms compete in fixed fees to attract shoppers.  We let 
$H$ represent the distribution of lump-sum fees.  Note that a 
lump-sum fee and a utility offered are linearly related through 
the following equation: $t(u) = v(0) - u$.   Therefore, lemmas 
\ref{lem:no_atom}-\ref{lem:no_gap} apply to characterize 
lump-sum fees offered in equilibrium.  In particular, these 
lemmas imply that $H$ has a compact support and is continuously 
increasing in its support.  Also the highest fixed-fee in the 
support must be no greater than the lowest of $v(0)$ and the 
reservation fee, which we define as follows.  

A reservation fee, denoted by $t_R$, is the highest lump-sum fee 
at which nonshoppers buy outright, thus terminating search.  To 
determine the reservation fee we equate the expected benefit of 
search to the cost of search when $t_R$ is the lowest price a 
nonshopper has observed so far:
\begin{equation}\label{eq:tR}
	\int^{t_R}_{\underline{t}}H(t)dt = s.
\end{equation}
Here $\underline{t}$ is the lowest lump-sum fee in the support 
of $H$.  Notice that the expected benefit of search, given by 
the left-hand side of the equation, strictly increases with 
$t_R$.  Then, there exists a unique reservation fee that solves 
the equation if a solution exists.  A solution clearly exists 
for small search costs.  If a solution does not exist we set 
$t_R = v(0)$.  

While two-part tariffs---with their linear prices equal to the 
production marginal cost and lump-sum fees distributed according 
to $H$---fully characterize firms optimal strategies, the 
reservation fee characterizes nonshoppers' optimal search 
strategy.  The equilibrium so characterized exists if the firms' 
(nonlinear) pricing policies and nonshoppers' optimal search 
rule are consistent.  This is indeed the case as shown in the 
next proposition.

\begin{proposition}\label{prop:seq_nonlin}
	For any $\lambda\in (0,1)$ and $s >0$, there exists a unique 
	symmetric RPE given by $(p,H, t_R)$.  The equilibrium two- 
	part tariffs are given by $p=0$ and $H$, where
	\begin{equation}\label{eq:H}
		H(t) = 1 - \left[\frac{1-\lambda}{n \lambda} 
		\left(\frac{\overline{t}}{t}-1\right) 
		\right]^{\frac{1}{n-1}} \ \mbox{with support} \ 
		\left[\underline{t}, \overline{t}\right]
	\end{equation}
	and $\overline{t} = \min\{t_R,v(0)\}$. Furthermore, there 
	exists $\overline{s}>0$ such that $t_R$ is determined by 
	\eqref{eq:tR} for $s\leq \overline{s}$. 
\end{proposition}

The proof is in the appendix and the intuition is as follows.  
To determine $H$ for any given $t_R$ we use the fact that an 
individual firm is indifferent of setting any fixed-fee in the 
support of $H$ and prefer them to fixed-fees outside the 
support.  (Note that $\underline{t}$ solves 
$H(\underline{t})=0$.) This distribution of fees is then 
consistent with nonshoppers' optimal search rule in 
\eqref{eq:tR} if the expected benefit of search increases with 
$t_R$ given that $H$ is a function of $t_R$ too.  We show in the 
appendix that this is indeed the case.

\subsection{Linear Prices}

We now turn our attention to the case where firm can offer only 
linear prices, as in \cite{stahl1989}.  Since the model with 
linear prices is a special case of a model with two-part tariffs 
where lump-sum fees are set equal to zero, similar arguments 
from the previous subsection apply to determine an 
equilibrium.   To avoid repetition we summarize the equilibrium 
result with linear prices as follows. 

There exists a unique symmetric equilibrium where firms choose a 
random per-consumer revenue according to $F$, determined by 
\begin{equation}\label{eq:F}
	F(\pi) = 1 - \left[\frac{1-\lambda}{n \lambda} 
	\left(\frac{\overline{\pi}}{\pi} - 
	1\right)\right]^{\frac{1}{n-1}}, \ \mbox{with support} \ 
	\left[\underline{\pi}, \overline{\pi}\right].
\end{equation}
Here $\underline{\pi}$ solves $F(\underline{\pi}) =0$ and 
$\overline{\pi} = \min\left\{\pi^m, \pi_R\right\}$ where $\pi_R$ 
is a reservation revenue which is uniquely determined by 
\begin{equation}\label{eq:piR}
	\int_{\underline{\pi}}^{\pi_R}(-v'(\pi))F(\pi)d\pi 
	=s
\end{equation}
if a solution exists.  A solution definitely exists for 
sufficiently small $s$ as the left-hand side of the equation, 
which represents the benefit of search, is positive and 
increasing in $\pi_R> 0$.  With a slight abuse of notation we 
let $\overline{s}>0$ be the cutoff value of the search cost such 
that the solution to \eqref{eq:piR} exists for $s\leq 
\overline{s}$.  If a solution does not exist we set $\pi_R= 
\pi^m$.  The equilibrium firm profit is then $(1-\lambda) 
\overline{\pi}/n$.

\begin{proposition}[Proposition 1 in \cite{stahl1989}]
	For any $0<\lambda<1$ and $n\geq 2$, there exists a unique 
	symmetric RPE given by $(F,\pi_R)$ where $F$ is 
	given by \eqref{eq:F} and $\pi_R$ is determined by 
	\eqref{eq:piR} for $s\leq \overline{s}$ and set equal to 
	$\pi^m$ otherwise.
\end{proposition}

For the proof we refer to \cite{stahl1989}, although a reader 
can easily establish the proof by following a line of argument 
presented for the case of nonlinear prices.  The intuition is 
similar to one behind Proposition \ref{prop:seq_nonlin} and thus 
we do not discuss it.

\section{Welfare Analysis}\label{s:welfare}

Our main aim is to compare the equilibrium market outcomes with 
nonlinear prices to those with linear prices.  The following 
proposition states the result.

\begin{proposition}\label{prop:welfare}
	Total surplus and firm profit are higher, whereas consumers 
	surplus is lower with nonlinear prices than with linear 
	prices.
\end{proposition}

The reasoning behind the first part follows from our 
equilibrium analysis.  Recall that the equilibrium production is 
socially optimal with nonlinear prices as the equilibrium linear 
price of two-part tariffs equals to the production marginal 
cost.  If firms compete in linear prices only, they choose 
positive levels of per-consumer revenue in equilibrium, which
means that  the equilibrium prices are bounded above the 
production marginal cost.  Therefore production is lower with 
linear prices than with nonlinear prices.

The proof of the last two parts of the proposition is in the 
appendix.  The main intuition lies in the \textit{effective} 
per-consumer demand firms face under the two different pricing 
schemes.  We know that with linear prices firms price at the 
elastic part of the demand curve in equilibrium.  One can think 
that with nonlinear prices firms face a unit (i.e., inelastic) 
demand for a hypothetical product which contains $q(0)$ amount 
of the real product.  Firms' market power is higher with 
inelastic demand than with elastic demand.  Therefore nonlinear 
prices allow sellers to soften competition by transforming the 
effective demand. This in turn raises profits and harms buyers.

\section{Extensions}\label{s:search}
 
In this section we demonstrate that our main result holds in 
different versions of our model.  We consider two extensions.  
In one of them, a buyer's search cost is a random draw from a 
distribution function  that is continuously increasing in its 
convex support (e.g., \cite{stahl1996}).  In the other 
extension we assume homogeneous search cost but heterogeneity 
in success of observing offers.  Specifically a searching buyer 
obtains offers from an unknown number of firms (e.g., 
\cite{wilde1977}, \cite{burdettjudd1983}).  This search 
protocol, known as \textit{noisy search}, equips sequential 
search with some properties of nonsequential search.

\subsection{Continuous Search Cost 
Distribution}\label{ss:continuous}

We first examine a case where each buyer's search cost is a 
random draw from interval $[0, \overline{c}]$ according to 
log-concave distribution function $G$ with density $g$.  We 
assume that $\overline{c}$ is positive and allow it to be 
infinite.  

\cite{stahl1996} shows that there exist a continuum of 
symmetric equilibria in linear prices.  In all these equilibria 
consumers do not search beyond the first firm and all firms 
charge the same price.  In terms of per-consumer revenue, any 
revenue in interval $[\pi^*,\pi^m]$ can be supported in 
equilibrium, where $\pi^*$ solves
\begin{equation*}
	\pi^* = \arg \max_{\pi} \left(1 - G\left[v(\pi^*) - 
	v(\pi)\right]\right) \pi.
\end{equation*}
Log-concavity of $G$ ensures a unique solution. The reason why 
$\pi^*$ can be supported in equilibrium is as follows. If all 
sellers charge a price to extract $\pi^*$, an individual 
deviation to a lower price is clearly unprofitable.  Consider an 
individual deviation to a slightly higher price that extracts 
revenue, say, $\widetilde{\pi}$.   This deviation results in a 
loss of consumers, as consumers with search costs lower than 
$v(\pi^*) - v(\widetilde{\pi})$ walk away to rival sellers. This 
loss of consumers prevents firms to individually raise their 
prices.   

The understanding why monopoly price is the same as that of 
Diamond-paradox.  If all sellers are expected to charge the 
monopoly price, searching an additional firm is not profitable.  
If however buyers do not search beyond the first firm, it is 
optimal for firms to offer the monopoly price, which justifies 
buyers' above expectation.

We now turn our attention to nonlinear prices.  As in the case 
of linear prices, we can apply \cite{stahl1996}'s argument to 
solve for equilibria with nonlinear prices.  One can easily 
check that analogous to equilibria with linear prices, buyers do 
not search beyond the first firm and firms play symmetric pure 
strategy (nonlinear) pricing in any equilibrium.  Moreover, 
optimal nonlinear price is a two-part tariff.  The linear price 
of a two-part tariff in any equilibrium is equal to the 
production marginal cost, which is implied by Lemma 
\ref{lem:p=0}.  Any fixed fee in interval $[t^*, v(0)]$ can be 
supported in equilibrium, where $t^*$ is determined as
\begin{equation*}
	t^* = \arg \max_{t} \left(1 - G\left[t - t^* \right]\right) 
	t.
\end{equation*}
The equilibrium industry profit then lies in $[t^*, 
v(0)]$ , while the consumer surplus lies in $[v(0) - t^*,0]$.

To compare market outcomes under different pricing regimes one 
needs to know which equilibria are played.  Selection of 
equilibria is oftentimes either based on refinement rule or 
one's judgment of which equilibrium fits real-world markets.  
However, it seems natural to us to focus on the most competitive 
equilibria, a route which is also taken in the existing 
literature.%
\footnote{\cite{janssenreshidi2020}, for instance, focus on
	equilibrium with linear prices generating revenue $\pi^*$ to 
	study price discrimination of retailer sellers by a 
	manufacturer.}
We state the main result in the following proposition.

\begin{proposition}\label{prop:seq_cont}
	Consider equilibria characterized by $\pi^*$ for linear 
	prices and $t^*$ for nonlinear prices.  With nonlinear 	
	prices sellers' profits and social welfare are higher, but 
	consumer surplus is lower than with linear prices. 
\end{proposition}

The reasoning behind the first two parts is fairly simple.  We 
can easily solve for $\pi^*$ and $t^*$ to obtain
\begin{equation*}
	\begin{aligned}
		\pi^* (-v'(\pi^*)) &&=&&& \frac{1}{g(0)},\\
		t^* &&=&&& \frac{1}{g(0)}.
	\end{aligned}
\end{equation*}
As in equilibrium with linear prices firms always choose prices 
at the elastic part of the demand, it must be that  
$-v'(\pi)>1$ for all $\pi\leq \pi^m$.  From this it directly 
follows that $0<\pi^* < t^*$ for any $0<g(0)<\infty$.  Note next 
that the linear price of equilibrium two-part tariffs equals to 
the production marginal cost, whereas the equilibrium prices are 
higher than the production marginal cost when firms can offer 
only linear prices.  Therefore the social welfare is higher with 
two-part tariffs than with linear prices.  

We prove in the appendix that consumers are better-off with 
linear prices than with nonlinear prices.  To understand the 
intuition it is useful to note that if the density of consumers 
with zero search cost $g(0)$ equals to $1/v(0)$, meaning that 
$t^* = v(0)$, the consumer surplus is zero with nonlinear prices 
but positive with linear prices.  As the density of consumers 
with zero search cost increases without bounds, the equilibrium 
profits converge to zero in both pricing regimes.  Therefore 
buyers receive the entire social surplus in both pricing regimes 
when the density of consumers with zero search cost becomes very 
large.   The challenge is then to show that the consumer surplus 
is increasing in the density of consumers with zero search cost 
more slowly with linear price than with nonlinear prices.  We 
prove in the appendix that this is the case.

\subsection{Noisy Search}\label{ss:noisy}

We finally turn our attention to markets where buyers employ 
noisy search, which incorporates some properties of 
nonsequential search into the model of sequential search.  Key 
features of nonsequential search is that a buyer chooses a 
number of firms to search after which the search is terminated.  
Noisy search allows such a consumer to search again if she is 
not satisfied with offers she received, yet it abstracts from 
formally modeling nonsequential search.  Therefore with noisy 
search a buyer requests deals from (searches) $m\geq 2$ number 
of firms at the beginning of each search round and receives 
responses from an unknown number of them, say $k$, at the end of 
a search round.  Here $k$ is a random variable with $1\leq 
k\leq m$.  At the end of a search round, a buyer decides 
whether to terminate the search or to send requests to $m$ more 
firms in the next search round.  If the search is terminated the 
buyer can either make a purchase at the lowest observed price or 
drop out of the market.  With a slight abuse of notation we let 
$s>0$ represent the cost of searching $m$ firms in each search 
round.

Noisy search is prevalent in markets where buyers observe search 
results with a delay.  For instance, in mortgage markets a 
consumer sends applications to several banks and receives 
replies from them after some time.  Some of those banks may not 
quote a price because they find the consumer too risky to
cooperate with.  In addition, before contacting banks, the 
consumer faces uncertainty about which banks will approve her 
application.%
\footnote{\cite{atayev2019a} analyzes uncertain product 
availability in search markets.}  

\subsubsection{Additional Assumptions}

To keep the model tractable and ensure uniqueness of 
equilibrium, we introduce some additional assumption.  Our first 
assumption is that there are infinitely many firms.  This 
assumptions ensures that buyers search rule is stationary.  
Search strategy of buyers may be nonstationary with finite 
number of firms.  This is because with finite number of firms 
the expected benefit of search is declining.  For instance, 
suppose that there are ten firms, buyers search four firms in 
each search round ($m=4$) and a buyer has observed two 
firms' offers in the first search round.  The expected benefit 
of searching in the second round equals to that of searching in 
the third search round, as in each of those rounds a buyer may 
receive up to four new offers.  To contrast, assume that the 
buyer received four offers in the first search round.  Then, the 
expected benefit of searching in the second round may be higher 
than that searching in the third search round.  This is because 
a buyer may receive four new offers in the second search round 
in which case she can receive at most two offers in the third 
search round.

Our next assumption is that a strictly positive (equal) share 
of buyers observe offers of each firm at the first search 
round.  We also let $\mu(k)$ represent the probability with 
which a buyer receives $k(\leq m)$ number of responses, where 
$\mu(0)=0$.  Note that if $\mu(1)=1$ the unique equilibrium is 
Diamond-paradox.  Conversely, if $\mu(1)=0$ the unique 
equilibrium is a textbook Bertrand equilibrium.  To rule out 
these trivial equilibria we assume that $0< \mu(1)<1$, and to 
ensure uniqueness of equilibrium we set $0<\mu(2)<1$.

The game unfolds as follows.  First, firms simultaneously choose 
their offers.  Second, without knowing offers, buyers engage in 
the first round of noisy search.  After observing offers, buyers 
may make purchase, drop out of the market, or search in the 
second search round.  The cycle continues until all buyers 
either make a purchase or drop out of the market.  The solution 
concept is an RPE.

\subsubsection{Analysis}\label{ss:eq_noisy}

With only linear prices allowed our model boils down to the 
model of \cite{burdettjudd1983}.  The authors show existence of 
a unique equilibrium which is characterized by a symmetric 
mixed-strategy pricing.  In terms of per-consumer revenue, firms 
randomly draw a revenue $\pi$ according to distribution $F$, 
which is strictly increasing in its support and solves
\begin{equation}\label{eq:F_noisy}
	\sum_{k=1}^m k \mu(k) (1-F(\pi))^{k-1} \pi = \mu(1) 
	\overline{\pi}
\end{equation}
and has support $[\underline{\pi}, \overline{\pi}]$. Here 
$\underline{\pi}$ solves $F(\underline{\pi})=0$ and 
$\overline{\pi} = \min\{\pi_R, \pi^m\}$, where $\pi_R$ is 
implicitly determined by
\begin{equation}\label{eq:pir_noisy}
	\int_{\underline{\pi}}^{\pi_R} (-v'(\pi))\sum_{k=1}^{m} 
	\mu(k) (1-F(\pi))^{k-1} d \pi = s.
\end{equation}
If a solution to the equation does not exist, we set 
$\pi_R=\pi^m$. 

Some observations are in order.  As in our main model, 
buyers do not search beyond the first search round in 
equilibrium.  We can hence consider buyers who observe only one 
price as nonshoppers.  Also the equilibrium prices are dispersed 
and the distribution of per-consumer revenue is strictly and 
continuously increasing in its compact support.  The only 
difference is that the equilibrium is unique with noisy search, 
whereas with the sequential search there exist asymmetric 
equilibria for $n\geq 3$.%
\footnote{\cite{bayeetal1992} show existence of asymmetric 
	equilibria where the share of buyers who observe exactly two
	prices is zero, and \cite{johnenronayne2020} show uniqueness 
	of symmetric equilibrium where that share of buyers is 
	positive.}%   

Next, the equilibrium with noisy search also resembles that with 
nonsequential search.  Particularly, the search is terminated 
after the first search round, which is exogenously imposed in 
non-sequential search markets.  There is however an important 
difference.  In nonsequential search models the highest price 
in the support of equilibrium price distribution equals to the 
monopoly price.  This highest price does not depend on the 
search cost.  In our model with noisy search, the search cost is 
a key determinant of the highest price in the support of the 
equilibrium price distribution.

The equilibrium with nonlinear prices is also similar to the one 
in our main model.   To avoid repetition we state the 
result outright.  In equilibrium firms offer two-part tariffs 
with $p=0$ and $t$ distributed according to $H$ that is strictly 
increasing in its support and solves
\begin{equation}\label{eq:H_noisy}
\sum_{k=1}^m k \mu(k) (1-H(t))^{k-1} t = \mu(1) t_R.
\end{equation}
The support of $H$ is $[\underline{t}, t_R]$ which are 
determined by $H(\underline{t})=0$ and 
\begin{equation}\label{eq:tr_noisy}
	\sum_{k=1}^{m} \mu(k) \int_{\underline{t}}^{t_R} 
	(1-H(t))^{k-1} d t=s.
\end{equation}
If a solution to the equation does not exist or $t_R>v(0)$, we 
set $t_R=v(0)$.

We now turn to welfare analysis. As the equilibria under the two 
pricing regimes with noisy search are qualitatively the same as 
the respective symmetric equilibria with sequential search in 
our main model, the welfare analysis with noisy search is 
qualitatively the same as that with the sequential search.  
Specifically, we have the following result.

\begin{proposition}\label{prop:welfare_noisy}
	With noisy search, social welfare and firm profits are 
	higher, whereas consumer surplus is lower with nonlinear 
	price than with linear prices.
\end{proposition}

The reasoning and the intuition behind this result is similar to 
one behind Proposition \ref{prop:welfare}.  Therefore we 
omit them.

\section{Conclusion}

We see our paper as the first one to examine welfare effects of 
nonlinear prices in homogeneous goods markets with search.  We 
demonstrate that traditional models which consider linear price 
understate firms' market power.  With nonlinear prices available 
in their hands, sellers can make effective per-buyer demand 
less elastic and soften competition.

\newpage

\appendix

\begin{singlespace}

\section{Proofs}

\subsection{Proof of Proposition \ref{prop:seq_nonlin}}

We start with existence.  Consider firms pricing strategy for 
any given $t_R$.  Suppose that an individual firm deviates to 
linear pricing strategy and the deviation price is $p_d$. The 
expected profit of the deviating firm equals to
\begin{equation*}
	\left(\frac{1-\lambda}{n} + \lambda \left[1 -  
	H\left(\int_{0}^{p_d} q(z) dz\right) \right]^{n-1} \right) 
	q(p_d)p_d = \frac{1-\lambda}{n} 	
	\left(\frac{q(p)p}{\int_{0}^{p}q(z)dz} \right) 
	\overline{t},
\end{equation*}
where we obtained the equality using \eqref{eq:H} and recall 
that $\overline{t} = \min\{t_R,v(0)\}$. The optimal $p_d$ either 
solves 
\begin{equation*}
	\left.\frac{\partial}{\partial p}\left( 
	\frac{q(p)p}{\int_{0}^{p}q(z)dz} \right) \right|_{p=p_d} =0
\end{equation*}
or satisfies $H(\int_{0}^{p_d}q(z)dz)=0$.  In the 
former case, the equation determining $p_d$ is 
\begin{equation}\label{eq:foc_pd}
	q'(p_d)p_d + q(p_d) - \frac{q(p_d)^2 p_d}{\int_{0}^{p_d} 
		q(z) dz} = 0.
\end{equation}
As the denominator of the third term on the LHS of the 
equation are positive,  it must be $p_d<p^m$ since $p^m$ 
solves $q'(p)p + q(p)=0$.  However, a firm can offer a two-part 
tariff $(p=0, t = \int_{0}^{p_d} q(z)dz)$ that yields the same 
utility to buyers are the deviation price $p_d$ and increase its 
profit, as $\int_{0}^{p_d}q(z)dz > q(p_d)p_d$. In the case where 
$p_d$ satisfies $H(\int_{0}^{p_d} q(z) dz) = 0$, we have 
$\int^{\overline{p}}_{p_d}q(z)dz = \underline{t}$.  As in the 
previous case, offering the two-part tariff that gives 
buyers utility $\int_{0}^{\overline{p}}q(z)dz - \underline{t}$ 
yields a higher expected profit than offering linear price 
$p_d$, as per-consumer revenue with $p_d$ equals to 
$q(p_d)p_d$ and that with the two-part 
tariff equals to $\underline{t}$ where $\underline{t} = 
\int_{0}^{p_d}q(z)dz > q(p_d)p_d$.  In either case, we arrive at 
a contradiction.  This means that there is no profitable 
deviation to 
linear pricing.  

To complete the proof of existence it is left to show that, 
given $H$, there exists a unique $t_R$ that solves 
\eqref{eq:tR} for $s\leq \overline{s}$. To do that we need to 
show that the LHS of \eqref{eq:tR}, which is positive, 
strictly increases with $t_R$.  The derivative of the LHS of 
\eqref{eq:tR} w.r.t. $t_R$ is $1 - \int_{\underline{t}}^{t_R} 
\partial H(t)/\partial t_R dt$.  As
\begin{equation*}
	\frac{\partial H(t)}{\partial t_R} = 
	\frac{t}{t_R} \times \frac{\partial H(t)}{\partial t} \leq  
	\frac{\partial H(t)}{\partial t} \ \ \mbox{for any} 
	\ t\leq t_R,
\end{equation*}
we have $1 - \int_{\underline{t}}^{t_R} 
\partial H(t)/\partial t_R dt > 1 - \int_{\underline{t}}^{t_R} 
\partial H(t)/\partial t dt =0$.  This means that the LHS of 
\eqref{eq:tR} is indeed strictly increasing in $t_R$.  Then, 
there must exist a unique $t_R$ solving \eqref{eq:tR} for 
$s\leq \overline{s}$.

We next prove that this is a unique symmetric equilibrium. 
Clearly, there cannot be other symmetric equilibria in two-part 
tariffs as $t_R$ and $H$ are uniquely determined by 
\eqref{eq:tR} and \eqref{eq:H}, respectively, for $p=0$.  
Then, it suffices to show that there does not exist another 
symmetric equilibrium in linear prices.  For contradiction, 
assume that there exists such an equilibrium with some price 
$\widetilde{p}$ in the support of equilibrium price distribution 
(in the there is a pure-strategy symmetric equilibrium 
$\widetilde{p}$ is the equilibrium price) and the per-consumer 
revenue $q(\widetilde{p}) \widetilde{p} /n$.  However, the firm 
can offer a two-part tariff with $p=0$ and $t = 
\int_{0}^{\widetilde{p}}q(z)dz$ which would yield the same 
utility to consumers but improve the firm's per-consumer 
revenue.  The last point can be seen as follows: 
$\int_{0}^{\widetilde{p}}q(z)dz > q(\widetilde{p}) 
\widetilde{p}$.  This contradicts to our assumption that there 
exists a symmetric equilibrium in linear prices. This 
establishes uniqueness of the symmetric equilibrium with 
two-part tariffs.

The proof of the proposition is now complete.

\subsection{Proof of Proposition \ref{prop:welfare}}

For the proof it suffices to compare profits and consumer 
surpluses under two pricing regimes.  The industry profit with 
nonlinear prices, which is 	$(1-\lambda) \overline{t}$, is 
higher than that with linear prices, which is $(1-\lambda) 
\overline{\pi}$, if	$\overline{t}>\overline{\pi}$. To see that 
this is the case, first recall that $F$ and $H$ are similar and 
the only difference is their supports.  Using formulas which 
determine these distributions, observe that 	
$\underline{\pi}/\overline{\pi} = (1-\lambda)/(1 + (n-1) 
\lambda)$ and $\underline{t}/\overline{t} = (1-\lambda)/(1 + 
(n-1)\lambda)$, which implies 
\begin{equation}\label{eq:pi0/piM}
	\frac{\overline{\pi}} {\underline{\pi}} = 
	\frac{\overline{t}} {\underline{t}}.
\end{equation}

Second, recall that the upper bound of $F$ is determined by 	
\eqref{eq:piR} and that of $H$ is determined by 
\eqref{eq:tR} for small $c$ (for high $c$ we know that 
$\overline{\pi}<\overline{t}$ from Proposition 
\ref{prop:welfare}).  As firms price at the elastic part of the 
demand meaning that $-v'(\pi)>1$ for all $\pi \leq \pi^m$ in 	
\eqref{eq:piR}, it must be that $\overline{\pi} -\underline{\pi} 
< \overline{t}- \underline{t}$.  Divide both sides of the 
inequality by $\overline{\pi} 	\overline{t}$ to obtain 
\begin{equation*}
	\frac{1}{\overline{t}} - \frac{1-\lambda}{1  + (n-1)  
		\lambda} \times \frac{1}{\overline{t}} < 
	\frac{1}{\overline{\pi}} - 	\frac{1-\lambda}{1 + (n-1) 
		\lambda} \times \frac{1}{\overline{\pi}},
\end{equation*}
which can be reduced to $\overline{t}>\overline{\pi}$, which was 
required to prove.

To compare consumer surplus under different pricing regimes 	
we start noting that the consumer surplus generating the 	
equilibrium profit with linear prices is not less than 
\begin{equation*}
	\frac{n-1}{n} (1-\lambda) v(\overline{\pi}) + \left(1 - 
	\frac{n-1}{n} (1-\lambda)\right) v(\underline{\pi}).
\end{equation*}
With two-part tariffs the consumer surplus equals to
\begin{equation*}
	\lambda v(0) + (1-\lambda) \left(\int_{0}^{\overline{p}} 
	q(z)dz - \overline{t}\right).
\end{equation*}
The former is greater than the latter if
\begin{equation*}
	v(\underline{\pi}) \geq \frac{\lambda}{1 - \frac{n-1}{n} 
		(1-\lambda)} v(0) + \frac{1-\lambda}{1 - \frac{n-1}{n} 
		(1-\lambda)} \left(v(0) - \overline{t}\right) - 	
	\frac{\frac{n-1}{n} (1-\lambda)}{1 - \frac{n-1}{n} 
		(1-\lambda)} v(\overline{\pi}).
\end{equation*}
Since $2(n-1)/n\geq 1$ for any $n\geq 2$, the inequality 
certainly holds if
\begin{equation}\label{eq:cs_seq}
	v(\underline{\pi}) \geq  \frac{\lambda}{1 - \frac{n-1}{n} 
		(1-\lambda)} v(0) + \frac{\frac{n-1}{n} (1-\lambda)}{1 - 
		\frac{n-1}{n} (1-\lambda)} v(\overline{\pi}) - 
	\frac{1-\lambda}{1 - \frac{n-1}{n} (1-\lambda)} 
	\left(v(\overline{\pi}) - v(0) - \overline{t}\right).
\end{equation}
As $v(\cdot)$ is a concave function, the LHS of the inequality 
is greater than the sum of the first two-terms on the RHS.  
Then, the inequality certainly holds if $v(\overline{\pi}) \geq 
v(0) - \overline{t}$, which we prove as follows. Note that both 
sides of this inequality converge to $v(0)$ as $s \downarrow 
0$.  Since both sides of the inequality are decreasing in $s$, 
the inequality holds if the derivative of the LHS is less 
negative than that of the RHS.  The derivative of the LHS is 
$v'(\overline{\pi}) \partial \overline{\pi}/\partial s$ and that 
of the RHS is $-\partial \overline{t} / \partial s$. To derive 
these derivatives we differentiate \eqref{eq:tR} and 
\eqref{eq:piR} w.r.t. $s$ to obtain
\begin{equation*}
	\begin{aligned}
		\frac{\partial \overline{\pi}}{\partial s} \left(- 		
		v'(\overline{\pi}) - \frac{1}{n-1} 	
		\left(\frac{1-\lambda}{1 + (n-1) 
		\lambda}\right)^{\frac{1}{n-1}} 
		\int_{\underline{\pi}}^{\overline{\pi}} (-v'(\pi)) 
		\left(\frac{\overline{\pi}}{\pi} - 1 
		\right)^{-\frac{n-2}{n-1}} 
		\frac{d \pi}{\pi} \right) = 1,\\			
		\frac{\partial \overline{t}}{\partial s} \left(1 - 
		\frac{1}{n-1} 
		\left(\frac{1-\lambda}{1 + (n-1) 
		\lambda}\right)^{\frac{1}{n-1}} 
		\int_{\underline{t}}^{\overline{t}} 
		\left(\frac{\overline{t}}{t} 
		- 1 \right)^{-\frac{n-2}{n-1}} \frac{d t}{t} \right) = 1.
	\end{aligned}
\end{equation*}
Then the derivatives of the LHS and the RHS of our inequality 
are, respectively,
\begin{equation*}
	\begin{aligned}
		\frac{1}{- 1 + \frac{1}{n-1} \left(\frac{1-\lambda}{1 
		+ 	(n-1) \lambda}\right)^{\frac{1}{n-1}} 
			\int_{\underline{\pi}}^{\overline{\pi}} 
			\frac{v'(\pi)}{v'(\overline{\pi})} 
			\left(\frac{\overline{\pi}}{\pi} - 1 
			\right)^{-\frac{n-2}{n-1}} 
			\frac{d \pi}{\pi} },\\			
		\frac{1}{- 1 + \frac{1}{n-1} \left(\frac{1-\lambda}{1 + 
		(n-1) \lambda}\right)^{\frac{1}{n-1}} 
			\int_{\underline{t}}^{\overline{t}} 
			\left(\frac{\overline{t}}{t} - 1 
			\right)^{-\frac{n-2}{n-1}} 
			\frac{d t}{t} }.
	\end{aligned}
\end{equation*}
The former is less negative than the latter if
\begin{equation*}
	\int_{\underline{t}}^{\overline{t}} 
	\left(\frac{\overline{t}}{t} - 1 \right)^{-\frac{n-2}{n-1}} 
	\frac{d t}{t} \geq \int_{\underline{\pi}}^{\overline{\pi}} 
	\frac{v'(\pi)}{v'(\overline{\pi})} 	
	\left(\frac{\overline{\pi}}{\pi} - 1 
	\right)^{-\frac{n-2}{n-1}} 	\frac{d \pi}{\pi} ,
\end{equation*}
which is true as (i) $v'(\pi)<0$ and $v'(\cdot)$ is a decreasing 
function (recall concavity of $v(\cdot)$), meaning that 
$v'(\pi)/v'(\overline{\pi})<1$ for any $0\leq \pi \leq 
\overline{\pi}$, and (ii) we know from the first part of the 	
proof that $\overline{t} - \underline{t} \geq  \overline{\pi} - 
\underline{\pi}$, $\overline{t} \geq \overline{\pi}$ and 
$\underline{t} \geq \underline{\pi}$.  This proves that the 
derivative of $v(\overline{\pi})$ w.r.t. $s$ is less negative 
than the derivative of $v(0) - \overline{t}$ .  Then 
(recalling the limiting results when $s \downarrow 0$) it is 	
indeed true that $v(\overline{\pi}) \geq v(0) - \overline{t}$ 	
for any $s >0$.  

This inequality proves in its turn that the inequality in 	
\eqref{eq:cs_seq} is true.  Then, the consumer surplus is 	
indeed lower with nonlinear prices than with linear prices.  	
The proof of the proposition is now complete.

	\subsection{Proof of Proposition \ref{prop:seq_cont}}
	
	To compare the consumer surpluses, we first note that the 
	equilibrium consumer surplus with linear prices is always 
	positive.  We next note that the consumer surplus increases 
	with $g(0)$ in both pricing regimes.  Moreover, the consumer 
	surplus with nonlinear prices equals to zero if $g(0) = 
	1/v(0)$, whereas the consumer surplus under both pricing 
	regimes converges to $v(0)$ as $g(0) \to \infty$.  Then, 
	consumers are worse-off with nonlinear prices than with 
	linear prices if consumer surplus is increasing in 
	$g(0)(>1/v(0))$ more slowly with linear prices than with 
	nonlinear prices.  The condition holds if the derivative of 
	the consumers surplus with respect to $g(0)$ is lower with 
	linear prices than with nonlinear prices.  These derivatives 
	are respectively,
	\begin{equation*}
		\begin{aligned}
			\frac{d v(\pi^*)}{d g(0)} &&=&&& v'(\pi^*) \frac{d 
			\pi^*}{d g(0)},\\
			\frac{d (v(0)-t^*)}{d g(0)} &&=&&& - \frac{d t^*}{d 
			g(0)},
		\end{aligned}
	\end{equation*}
	where $v(\pi^*)$ is the equilibrium consumers surplus with 
	linear prices.

	To evaluate these derivatives, we differentiate the 
	equations solving for $\pi^*$ and $t^*$ with respect to 
	$g(0)$ to obtain
	\begin{equation*}
		\begin{aligned}
			\frac{d \pi^*}{d g(0)} &&=&&& \frac{1}{[g(0)]^2 
				\left(v'(\pi^*)  + v''(\pi^*) \pi^*\right)},\\
			\frac{d t^*}{d g(0)} &&=&&& - \frac{1}{[g(0)]^2}.
		\end{aligned}
	\end{equation*}
	Then, the derivative of the consumers surplus with linear 
	prices is lower than that with nonlinear prices if
	\begin{equation*}
		\frac{1}{1  + \frac{v''(\pi^*)}{v'(\pi^*)} \pi^*} \leq 1.
	\end{equation*}
	This inequality holds since $v''(\pi^*)/v'(\pi^*)>0$, which 
	follows from the fact that $v(\cdot)$ is a decreasing 
	concave function.  This shows that the consumer surplus is 
	rising more slowly in $g(0)$ with linear prices than with 
	nonlinear prices for all $g(0)>1/v(0)$.  This fact---along 
	with the two facts mentioned in the previous paragraph---
	proves that the consumer surplus is higher with linear 
	prices than with nonlinear prices.

 \newpage

\bibliographystyle{aer}
\bibliography{xx}{}

\end{singlespace}	

\ifx\undefined\bysame
\newcommand{\bysame}{\leavevmode\hbox 
to\leftmargin{\hrulefill\,\,}}
\fi

\end{document}